\documentclass[%
 reprint,
superscriptaddress,
 aps,prb,
]{revtex4-2}
\usepackage{graphicx}
\usepackage{dcolumn}
\usepackage{bm}
\usepackage{physics}
\usepackage{amsmath}
\usepackage{bbm}
\usepackage[version=3]{mhchem}
\usepackage{subfig}
\usepackage{color}
\newcommand{\vk}{\vb*{k}}

\renewcommand{\vr}{\vb*{r}}
\newcommand{\vlr}{\vb*{R}}
\newcommand{\mean}[1]{\left\langle #1  \right\rangle}

\newcommand{\ep}{\varepsilon}

\newcommand{\kbt}{k_B T}
\newcommand{\iwn}{i\omega_{n}}

\newcommand{\rkfwa}{\pqty{\vlr,\vk_{\rm F}^{\alpha};\iwn}}
\newcommand{\rkfwb}{\pqty{\vlr,\vk_{\rm F}^{\beta};\iwn}}

\begin{document}

\title{Theoretical analysis of anisotropic upper critical field of superconductivity in nodal-line semimetals}

\author{\normalsize Junya Endo}
\affiliation{Department of Physics, University of Tokyo, Bunkyo, Tokyo 113-0033, Japan}

\author{\normalsize Hiroyasu Matsuura}%
\affiliation{Department of Physics, University of Tokyo, Bunkyo, Tokyo 113-0033, Japan}

\author{\normalsize Masao Ogata}
\affiliation{Department of Physics, University of Tokyo, Bunkyo, Tokyo 113-0033, Japan}
\affiliation{Trans-scale Quantum Science Institute, University of Tokyo, Bunkyo-ku, Tokyo 113-0033, Japan}

\date{\today}

\begin{abstract}
    We study the properties of the upper critical field of superconductivity in nodal-line semimetals in a continuous model, which has a nodal-line on the $k_{z} = 0$ plane. Using the semiclassical Green's function method, we calculate the upper critical field for the two limiting cases: the dirty limit with many impurities and the clean limit with few impurities. The results show the large anisotropy of the magnitude of the upper critical field and the unusual temperature dependence. The obtained results are compared with recent experimental data of PbTaSe$_{2}$.
\end{abstract}

\maketitle

\section{Introduction}
In the standard systems, electron bands avoid band crossing because of band repulsion.
However, some symmetry protects the band crossing of the conduction and valence bands. Materials that exhibit such kind of non-accidental band crossings are called topological semimetals\cite{RevModPhys.82.3045}.
Three-dimensional topological semimetals are classified into two types according to the dimension of the crossing, i.e., whether the band crossing appears as a point (zero-dimensional) or as a line (one-dimensional).
The former includes gapless Dirac and Weyl semimetals. Unusual phenomena such as anomalous Hall effect and spin Hall effect that occur in these systems have been vigorously studied both theoretically and experimentally\cite{Shuichi_Murakami_2007,PhysRevB.83.205101,PhysRevLett.107.186806,PhysRevB.85.195320,doi:10.7566/JPSJ.83.063709,PhysRevLett.113.187202,Ado_2015,doi:10.1080/00018732.2015.1068524}. Some of these phenomena, especially those common to the Dirac and Weyl semimetals, are due to their linear dispersion near the Fermi surface.
On the other hand, semimetals with one-dimensional band crossing are called nodal-line semimetals\cite{PhysRevB.84.235126,PhysRevB.92.045108,Fang_2016,PhysRevB.93.201104,Hirayama2017,Takane2018}. Their band structure can be interpreted as Dirac points and Dirac cones aligned along a line\cite{PhysRevB.104.035113}. Therefore, some features of the Dirac semimetals are inherited in the nodal-line semimetals. In addition, the nodal-line can be of various shapes. Therefore, the nodal-line semimetals are expected to have more degrees of freedom and richer physics than the Dirac and Weyl semimetals. In fact, they are known to induce peculiar behaviors in orbital magnetism and thermoelectric effect, for instance\cite{PhysRevB.104.155202,PhysRevB.105.085406}.
\par
PbTaSe$_2$ and SnTaS$_2$ are proposed to be nodal-line semimetals\cite{Bian2016,PhysRevB.93.245130,PhysRevB.40.12111,Jin2019,PhysRevB.103.035133}, which exhibit s-wave superconductivity\cite{LE20201349,PhysRevB.100.064516,PhysRevB.97.184510,PhysRevB.93.020503,PhysRevB.93.060506}. 
Interestingly, the superconductivity of PbTaSe$_2$ shows a large anisotropy and a peculiar temperature dependence of the upper critical field as follows\cite{PhysRevB.93.054520}: (1) The upper critical field parallel to the $ab$ plane is several times larger than that parallel to the $c$ axis. (2) The upper critical field decreases linearly as the temperature increases when the field is parallel to the $c$ axis. (3) The critical field increases slowly near the superconducting transition temperature. SnTaSe$_2$ also exhibits similar properties to (1) and (3)\cite{PhysRevB.100.064516}.
It is known that the behavior of $H_{c2}$ is highly dependent on the structure of the Fermi surface. Therefore, we can expect unconventional behavior of the upper critical field originating from the peculiar shape of the Fermi surface of nodal-line semimetals.
\par
In this paper, we theoretically study the upper critical field of superconductivity in nodal-line semimetals. The quasi-classical Green's functions allow the Gor'kov equations to be rewritten in simpler forms, such as the Eilenberger equations and the Usadel equations\cite{Eilenberger1968,1969JETP...28.1200L,PhysRevLett.25.507}. These equations make it possible to obtain the upper critical field numerically\cite{PhysRevB.67.184515,PhysRevB.70.054509}.
Using these methods, we clarify the anisotropy and the temperature dependence of the upper critical field in the two cases: the dirty limit and the clean limit. We will show that the obtained results are consistent with the above experimental results (1)-(3).
\par
\begin{figure}[htbp]
    \begin{center}
        \includegraphics[keepaspectratio, width=0.95\linewidth]{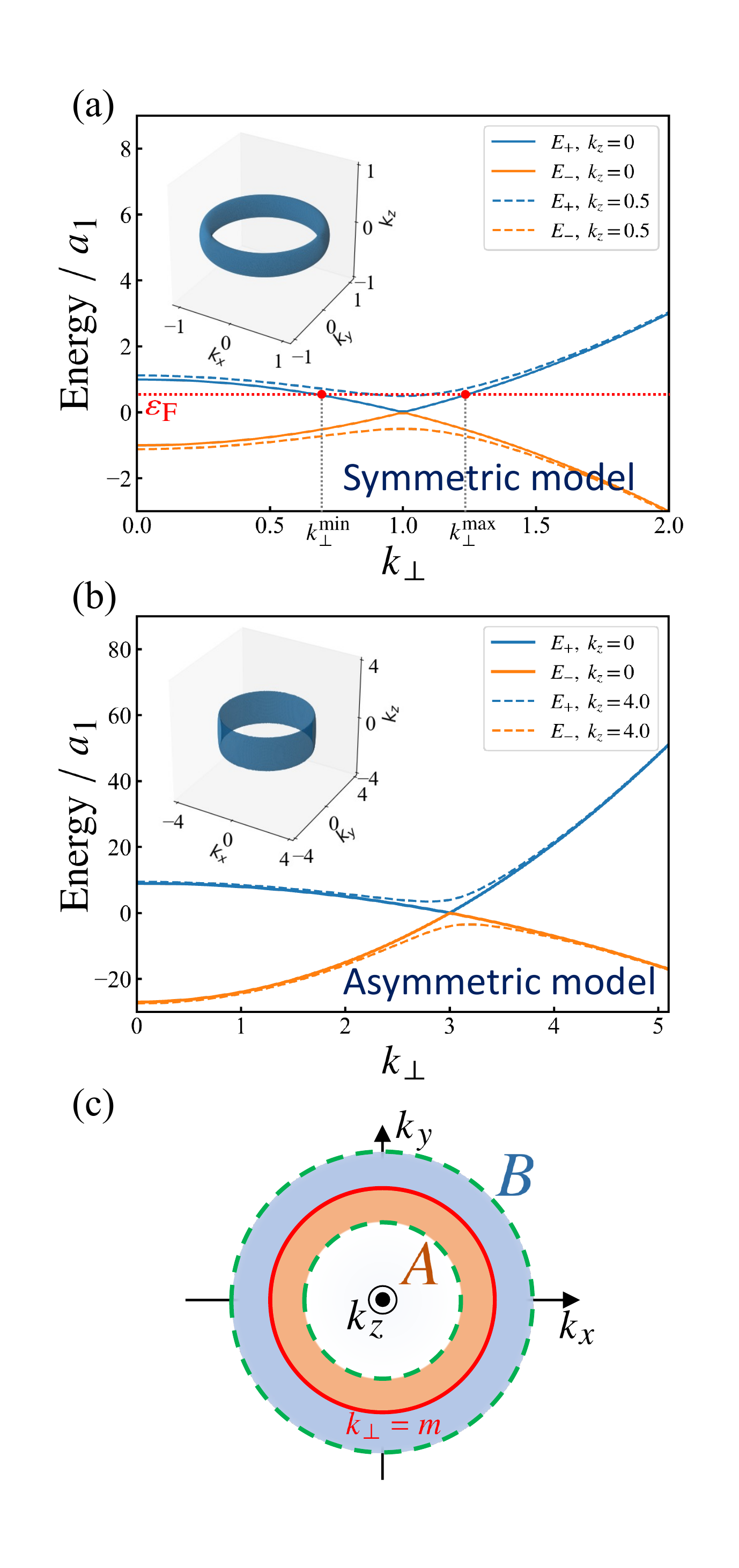}
        \caption{(a) Band dispersion of eq. \eqref{eq:dispersion} with $a_{2}/a_{1} = b/a_{1} = 1.0$ and $m=1.0$ in the cases of $k_{z} = 0$ and $k_{z} = 0.5$. The band crossing (nodal-line) appears when $k_{z} = 0$. An example of the Fermi energy level, $\ep_{\rm F}$ is shown by the red dotted line. When $k_{z}=0$, the Fermi surface is composed of two circles with radii $k_{\perp}=k_{\perp}^{\rm min}$ and $k_{\perp}^{\rm max}$, which are shown by the red circles. The inset shows the Fermi surface for $\ep_{\rm F}/a_{1} = 0.2$. (b) Band dispersion of with $a_{2}/a_{1} = 3.0$, $b/a_{1}=1.0$, and $m=3.0$ in the cases of $k_{z}=0$ and $k_{z}=4.0$. The inset shows the Fermi surface for $\ep_{\rm F}/a_1 = 1.0$. (c) Top view of the Fermi surface. Two branches $A$ and $B$ are divided by the red circle representing $k_{\perp}=m$ when $\abs{k_{z}}=\ep_{\rm F}/b$. $k_{\perp}$ satisfies $k_{\perp}^{\rm min} \leq k_{\perp} \leq m$ for the branch $A$ and $m \leq k_{\perp} \leq k_{\perp}^{\rm max}$ for the branch $B$. The green dotted lines correspond to $k_{\perp} = k_{\perp}^{\rm min}$ and $k_{\perp} = k_{\perp}^{\rm max}$.}
        \label{fig:dispersion}
    \end{center}
\end{figure}
\section{Model}
In real nodal-line semimetals, the nodal-lines can form some complicated shapes. However, in the following, we investigate a simple model of a circular nodal-line semimetal as a first step and find the peculiarity of the upper critical field ($H_{c2}$) in this model. We believe that this model sufficiently captures the essence of superconductivity in nodal-line semimetals.
\par
We use the following simple effective model Hamiltonian describing the nodal-line semimetal:  
\begin{eqnarray}
H = H_{0} + H^{\prime},
\end{eqnarray}
where $H_{0}$ is the kinetic energy Hamiltonian and $H^{\prime}$ is the two-body attractive interaction Hamiltonian.
We assume that $H_{0}$ is given by
\begin{equation}
    \label{eq:nodal_ham}
    H_{0}(\vk) = \mqty(a_{1}\pqty{k_{\perp}^{2} - m^{2}} & b k_{z} \\ b k_{z} & -a_{2}\pqty{k_{\perp}^{2} - m^{2}})~,
\end{equation}
where $k_{\perp}^{2} = k_{x}^{2} + k_{y}^{2}$, $k_{x}, k_{y},$ and $k_{z}$ represent dimensionless wave numbers, $a_{1}$, $a_{2}$, and $b$ are constants and $m$ is a dimensionless constant. In the following, we take $a_{1}$ as a unit of energy. 
Diagonalizing eq.\eqref{eq:nodal_ham}, the energy dispersion is obtained as 
\begin{equation}
    \begin{split}
        E_{\pm}(\vk) &= \frac{a_{1} - a_{2}}{2}(k_{\perp}^{2} - m^{2}) \\
        &~\pm \sqrt{\pqty{\frac{a_{1}+a_{2}}{2}}^{2}\pqty{k_{\perp}^{2} - m^{2}}^{2} + b^{2} k_{z}^{2}}. \label{eq:dispersion}
    \end{split}
\end{equation}
$E_+$ and $E_-$ bands intersect when $k_{z} = 0$, and their intersection forms a circular nodal-line in the $k_x$-$k_y$ plane of $k_{z} = 0$. For a certain Fermi energy $\ep_{\rm F}>0$ (the red dotted line of Fig. \ref{fig:dispersion}), the corresponding Fermi surface has a doughnut-like structure as shown in the inset of Fig. \ref{fig:dispersion}(a). In the $k_x$-$k_y$ ($k_{z}=0$), this Fermi surface is composed of two circles with radii $k_{\perp}=k_{\perp}^{\rm min}$ and $k_{\perp}^{\rm max}$, which are shown by the red circles in Fig. \ref{fig:dispersion}(a). ($k_{\perp}^{\rm min}=\sqrt{m^2 - \ep_{\rm F}/a_2}$ and $k_{\perp}^{\rm max}=\sqrt{m^2 + \ep_{\rm F}/a_1}$.) As $\abs{k_{z}}$ increases, the two circles shrink and finally merge at $k_{\perp}=m$ when $\abs{k_{z}}=\ep_{\rm F}/b$. 
Figure 1(b) shows the band dispersion at $k_{z} = 0, 4.0$ for $a_2/a_1 = 3.0$, $b/a_1 =1.0$, and $m = 3.0$. In this parameter set, we find the asymmetric band dispersion against the zero energy line. Therefore, we call the parameter settings for Fig. \ref{fig:dispersion}(a) (Fig. \ref{fig:dispersion}(b)) the symmetric (asymmetric) model.
\par
\section{Method}
Since the Fermi surface has a doughnut-like structure, and is composed of two circles for a fixed $k_{z}$, we divide the Fermi surface into two branches, $A$ and $B$, as shown in Fig. \ref{fig:dispersion}(c). The branch A is the part of the Fermi surface with $k_{\perp}^{\rm min} \leq k_{\perp} \leq m$ and the branch $B$ with $m \leq k_{\perp} \leq k_{\perp}^{\rm max}$. The two branches meet on the line $k_{\perp}=m$ when $\abs{k_{z}}=\ep_{\rm F}/b$. We assume that the superconductivity occurs in the very vicinity of each branch of the Fermi surface.  To take account of the electronic states on the Fermi surface, we use the semiclassical Green's functions defined by \cite{Eilenberger1968,kopnin2001theory}
\begin{equation}
    \begin{split}
        \check{g}^{\alpha}\rkfwa &= \mqty(\hat{g}^{\alpha}\rkfwa & \hat{f}^{\alpha}\rkfwa \\ \hat{f}^{\alpha\dagger}\rkfwa& \hat{g}^{\alpha}\rkfwa) \\ &= \int \frac{\dd \xi_{\vk}}{\pi} \check{\rho}_{z} \int \dd \tau \int \dd \vr ~ \mathrm{e}^{\iwn (\tau - \tau')}\mathrm{e}^{-i\vk_{\rm F}^{\alpha} \cdot \vr} \\ 
        & \hspace{-5mm} \times \mqty(\hat{G}^{\alpha}(\vr_{1},\tau; \vr_{2}, \tau') & \hat{F}^{\alpha}(\vr_{1},\tau; \vr_{2}, \tau') \\ \hat{F}^{\alpha\dagger}(\vr_{1},\tau; \vr_{2}, \tau') & \hat{\tilde{G}}^{\alpha}(\vr_{1},\tau; \vr_{2}, \tau')),
    \end{split}
    \label{eq:quasi_g}
\end{equation}
where $\alpha$ represents the branch $A$ or $B$, $\check{\rho}_{z} = \mathrm{diag}(\hat{\mathbbm{1}},-\hat{\mathbbm{1}})$, $\vlr = (\vr_{1} + \vr_{2})/2$ is the center of mass coordinate, $\vr = \vr_{1} - \vr_{2}$ is the relative coordinate, $\xi_{\vk} = \ep_{\vk} - \mu$ is the energy measured from the chemical potential, $\tau$ represents the imaginary time, and $\omega_{n} = (2n+1)\pi \kbt$ is the fermionic Matsubara frequency.
Here, $\hat{G}^{\alpha}$ and $\hat{F}^{\alpha}$ are the normal and anomalous Green's functions in the form of $2 \times 2$ matrix in the Nambu-Gor'kov space. In addition, $\tilde{G}$ is defined as $-\mean{\psi^{\dagger}\psi}$, similar to the normal Green's function.
The $4 \times 4$ matrix $\check{g}^{\alpha}$ is normalized as $(\check{g}^{\alpha})^2 = -\check{\mathbbm{1}}$. 
\par
Considering the equation of motion for $\hat{G}^{\alpha}$ and $\hat{F}^{\alpha}$ in the standard mean-field approximation, we can obtain the equation of motion for the quasi-classical Green's function $\check{g}^{\alpha}$\cite{Eilenberger1968,1969JETP...28.1200L}.
We assume that the coherence length is much larger than the Fermi wavelength near the upper critical field $H_{c2}$, and then the spatial variation of the gap function is small.
In fact, since $H_{c2} \sim 1000~\mathrm{Oe}$ and the lattice constant in PbTaSe$_2$ discussed in Section \ref{sec:analysis} is approximately $0.3~\mathrm{nm}$, the product of coherence length and Fermi wave number is estimated to be $k_{F}\xi \sim k_{F} \sqrt{\hbar \phi / 2\abs{e}H_{c2}} \sim 200 \gg 1$, which justifies our assumption.
Using this assumption, we obtain
\begin{equation}
    \begin{split}
        &\left[\pqty{i\omega_{n} + e\vb*{v}^{\alpha}_{\vk_{\rm F}^{\alpha}}\cdot \vb*{A}_{\vb*{R}}}\check{\rho}_{z} - \check{\Delta}^{\alpha}(\vb*{R},\vk_{\rm F}^{\alpha})-\check\Sigma^{\alpha}\pqty{\vk_{\rm F}^{\alpha},i\omega_{n}}\right. \\
        &\left.~,~\check{g}^{\alpha}\pqty{\vb*{R},\vk_{\rm F}^{\alpha},\iwn}\right]+ i\vb*{v}^{\alpha}_{\vk_{\rm F}}\cdot \nabla_{\vb*{R}} \check{g}^{\alpha}\pqty{\vb*{R},\vk_{\rm F}^{\alpha},\iwn} = 0,
    \end{split}
    \label{eq:Eilenberger}
\end{equation}
which is called Eilenberger equation\cite{Eilenberger1968}. Here, $\vb*{A}_{\vb*{R}}$ is the vector potential at $\vb*{R}$, $e<0$ is the electron charge, $\vb*{v}^{\alpha}_{\vk_{\rm F}^{\alpha}}$ is the Fermi velocity at $\vk_{\rm F}^{\alpha}$, and $[~,~]$ represents the commutation relation. The $4 \times 4$ matrix $\check{\Sigma}$ represents the self-energy, and $\check{\Delta}$ represents the gap function, which satisfies
\begin{equation}
    \check{\Delta}^{\alpha}\pqty{\vb*{R},\vk_{\rm F}^{\alpha}} = \mqty(0 & \hat{\Delta}^{\alpha}\pqty{\vb*{R},\vk_{\rm F}^{\alpha}} \\ \pqty{\hat{\Delta}^{\alpha}\pqty{\vb*{R},\vk_{\rm F}^{\alpha}}}^{*} & 0).
\end{equation}
The gap equation is
\begin{equation}
    \label{eq:dirty_gap}
    \hat{\Delta}^{\alpha} = 2\pi T \sum_{\omega_{n} > 0}^{\omega_{D}}\sum_{\beta}\lambda_{\alpha\beta}\mean{\hat{f}^{\beta}\rkfwb}_{\vb*{k}_{\rm F}^{\beta}},
\end{equation}
where $\omega_{D}$ is the Debye frequency, and $\mean{\cdots}_{\vk_{\rm F}^\alpha}$ represents the average over the Fermi surface of the corresponding branch. $\lambda_{\alpha\beta}$ are coupling constants defined by $N_{0}V_{\alpha \beta}$ where $N_{0}$ is the total density of states and $V_{\alpha \beta}$ is the attractive interaction between $\alpha$ and $\beta$ branches. The inter-branch coupling is also $(\alpha \neq \beta)$ included in this formalism. In the following, to estimate the upper critical field, $H_{c2}$, we introduce the two limits: the dirty limit and the clean limit. 
\par
For the case of dirty limit, assuming that the quasiclassical Green's function is almost isotropic and introducing the anisotropy by expanding to the first order of the Fermi velocity in eq. \eqref{eq:Eilenberger}, $\hat{f}^{\alpha}$ is given by solving the following equation called Usadel equation\cite{PhysRevLett.25.507}: 
\begin{equation}
    \label{eq:usadel}
    \pqty{2\omega_{n} - \sum_{\mu,\nu} D^{\alpha}_{\mu\nu}\vb*{\Pi}_{\mu}\vb*{\Pi}_{\nu}}\hat{f}^{\alpha}\rkfwa = 2\hat{\Delta}^{\alpha},
\end{equation}
where $D^{\alpha}_{\mu\nu} \propto \mean{v^{\alpha}_{\mu}v^{\alpha}_{\nu}}_{\vk_{\rm F}^{\alpha}}$ are the intra-band diffusivity tensors, $\vb*{\Pi} = \vb*{\nabla} + 2\pi i \vb*{A} / \phi$ is the covariant derivative operator, where $\phi$ is the magnetic flux quantum. 
We take the Landau gauge $\vb*{A} = H x_{\mu} \vb*{e}_{\nu} ~ (\mu \neq \nu)$, e.g., $\mu = x$ and $\nu = y$ if the magnetic field is oriented in the $z$ direction. Assuming $\hat{f}^{\alpha} \propto \hat{\Delta}^{\alpha}$, we obtain $\hat{f}^{\alpha} = \hat{\Delta}^{\alpha}/\pqty{\omega_{n} + K D^{\alpha}}$, with $K = \pi H / \phi$, and  $D^{\alpha} = \sqrt{D^{\alpha}_{\mu\mu} D^{\alpha}_{\nu\nu}}$. Here, we have assumed $D^{\alpha}_{\kappa\lambda} = 0$ when $\kappa \neq \lambda$. \par
As mentioned in the introduction, we study the upper critical field of s-wave superconductivity in the nodal-line semimetals. Therefore, we assume
\begin{equation}
    \hat{\Delta}^{\alpha} = \mqty(0 & \Delta^{\alpha} \\ -\Delta^{\alpha} & 0).
\end{equation}
Then, the gap equation eq. \eqref{eq:dirty_gap} together with eq. \eqref{eq:usadel} can be calculated as
\begin{equation}
    \Delta^{\alpha} = \sum_{\beta} \lambda_{\alpha \beta} \Delta^{\beta}\pqty{\log \frac{2\gamma \omega_{D}}{\pi T} - U\pqty{\frac{KD^{\beta}}{2\pi T}}}, \label{eq:gap2}
\end{equation}
where $U(x) = \Psi\pqty{x + 1/2} - \Psi\pqty{1/2}$, and $\Psi$ is the digamma function. In the above calculation, we have used the equation $\sum_{\omega_{n} > 0}^{\omega_{D}} 2\pi T/\pqty{\omega_{n} + X} = \log\pqty{2\gamma \omega_{D}/\pi T} - U\pqty{X/2\pi T}$ for $X > 0$. 
Using the matrix $P^{(d)}_{\alpha\beta} = \pqty{\log\pqty{2\gamma\omega_{D}/\pi T} - U\pqty{KD^{\beta}/2\pi T}}\lambda_{\alpha\beta}-\delta_{\alpha\beta}$, eq. \eqref{eq:gap2} becomes 
\begin{eqnarray}
P^{(d)}_{\alpha\beta}\Delta^{\beta} = 0.
\end{eqnarray}
Here, $(d)$ represents the dirty limit. To obtain a non-zero solution $\Delta^{\beta}$, the self-consistent equation is given by 
\begin{eqnarray}
\det P^{(d)} = 0. \label{eq:Sce}
\end{eqnarray}
In the dirty limit, $H_{c2}$ is obtained by solving eq. \eqref{eq:Sce}.
\par
Next, we discuss the case of clean limit. Assuming $\Sigma \ll k_{B}T_{c}$, eq. \eqref{eq:Eilenberger} becomes
\begin{equation}
    \begin{split}
        &\comm{\pqty{i\omega_{n} + e\vb*{v}_{\vk_{\rm F}^{\alpha}}\cdot \vb*{A}_{\vb*{R}}}\check{\rho}_{z} - \check{\Delta}^{\alpha}(\vb*{R},\vk_{\rm F}^{\alpha})}{~\check{g}^{\alpha}\pqty{\vb*{R},\vk_{\rm F}^{\alpha},\iwn}} \\
        &+ i\vb*{v}_{\vk_{\rm F}^{\alpha}}\cdot \nabla_{\vb*{R}} \check{g}^{\alpha}\pqty{\vb*{R},\vk_{\rm F}^{\alpha},\iwn} = 0.\
    \end{split}
\end{equation}
In order to estimate the upper critical field $H_{c2}$ efficiently for the clean limit, we use an approximation introduced by Brandt, Pesch, and Tewordt (BPT) \cite{BPT}, which is a method to calculate the free energy of multiband superconductivity. 
The BPT approximation consists of the following three approximations : (1) The spatial variation of $\Delta$ forms an Abrikosov lattice. (2) The magnetic flux density $\vb*{B}$ is spatially uniform. (3) The quasi-classical Green's function $\hat{g}^{\alpha}(\vb*{R})$ is spatially uniform. These assumptions are valid when the external magnetic field is near the upper critical field. 
Using the BPT approximation, the free energy of the multiband s-wave superconductor is generally given as
\begin{equation}
    \begin{split}
        \Omega &= \frac{\pqty{B - H}^2}{8\pi}  \\
        + &\sum_{\alpha \beta} \pqty{\rho^{(0),\alpha\beta} + \rho^{\alpha}(T,B) \delta_{\alpha\beta}} \pqty{\Delta^{\alpha}}^{*}\Delta^{\beta} + O\pqty{\abs{\Delta}^4},
    \end{split}
\end{equation}
with 
\begin{equation}
    \begin{split}
        \rho^{\alpha} (T,B) &= N^{\alpha} \left\langle \log\frac{T}{T_{c}} \right.\\ 
        + &\left. 2\pi T \sum_{n=0}^{\infty} \left\{ \frac{1}{\omega_{n}} + \frac{2\sqrt{\pi}\Lambda}{v^{\alpha}_{\perp}} \mathrm{W}\pqty{\frac{2i\omega_{n}\Lambda}{v^{\alpha}_{\perp}}}\right\} \right\rangle_{\vk_{\rm F}^{\alpha}},
    \end{split}
\end{equation}
where $\Lambda = \pqty{2\abs{e}B}^{-1/2}$, $v_{\perp}$ is the fermi velocity perpendicular to the magnetic field, $N^{\alpha}$ is the density of states of the $\alpha$th branch at the Fermi surface, and $\mathrm{W}(z) = \exp(-z^2)\mathrm{erfc}(-iz)$ is the Faddeeva function. $\rho^{(0),\alpha\beta} = \pqty{\lambda^{-1}}_{\alpha\beta} - N^{\alpha} \rho_{c} \delta_{\alpha\beta}$ is the coefficient of $\pqty{\Delta^{\alpha}}^{*}\Delta^{\beta}$ in the free energy when $B = 0$. Therefore, $\rho_{c}$ in $\rho^{(0),\alpha\beta}$ is determined from the condition $\det \rho^{(0)} = 0$.
\par
Since the gap function $\Delta$ is equal to zero at the critical point $H = H_{c2}$, the equation to obtain $H_{c2}$ by BPT is given by  
\begin{eqnarray}
\det P^{(c)} = 0,  \label{eq:Sce2}
\end{eqnarray}
where $P^{(c)}_{\alpha \beta} = \rho^{(0),\alpha\beta} + \rho^{\alpha}(T,B) \delta_{\alpha\beta}$.
\par
\section{Results}
\begin{figure}[t]
    \includegraphics[width=1.0\linewidth]{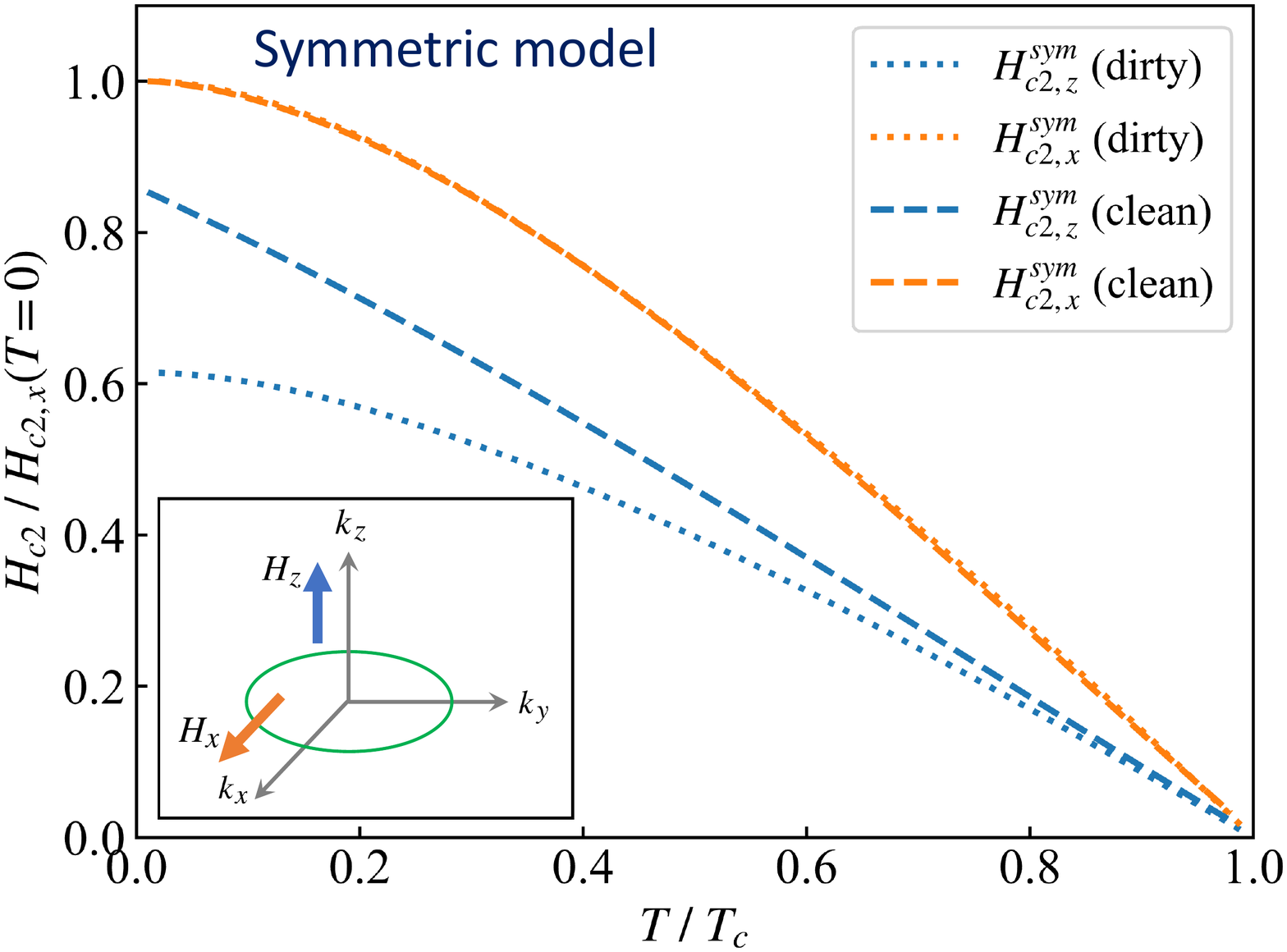}\\
    \caption{\label{fig:hc2_sym} Temperature dependence of the upper critical field for the symmetric model with $a_{1} = a_{2} = b = m = 1.0$. The behavior of $H_{c2,x}$ at the dirty and clean limits is very similar, and the lines in the graph overlap.} 
\end{figure}
\begin{figure}[t]
    \includegraphics[width=1.0\linewidth]{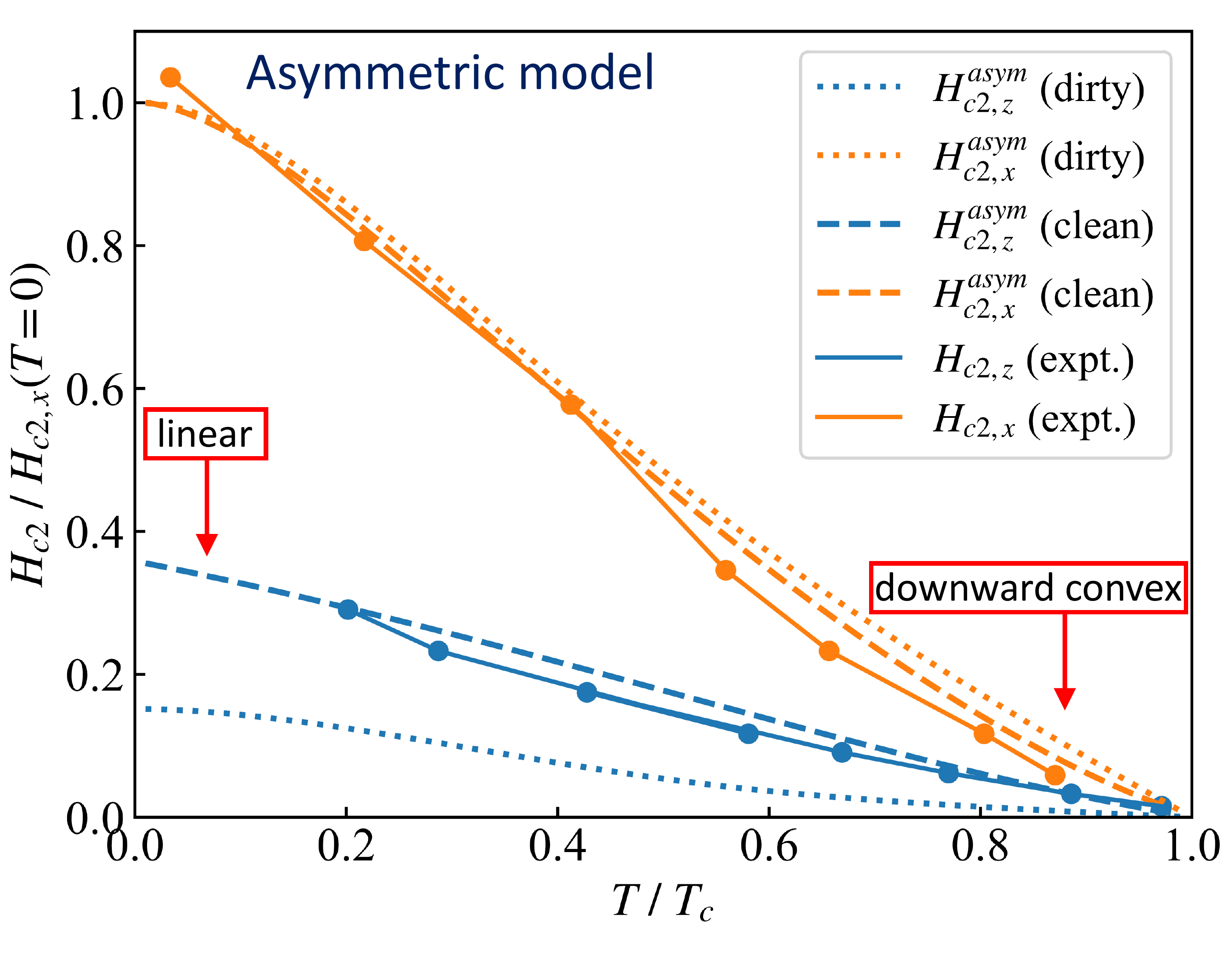}\\
    \caption{\label{fig:hc2_asym} Temperature dependence of the upper critical field for the asymmetric model. Experimental data on the upper critical field in PbTaSe$_2$\cite{PhysRevB.93.054520} are also included for comparison.} 
\end{figure}
Figure \ref{fig:hc2_sym} shows the temperature dependence of the upper critical field for the symmetric model with $a_{2}/a_{1} = b/a_{1} = 1.0$ and $m = 1.0$ in the dirty limit (dotted lines) and the clean limit (dashed lines). Here, we set the coupling constants as $\lambda_{AA}=1.0$, $\lambda_{BB}=0.8$, $\lambda_{BA}=0.2$, and $\lambda_{AB}=0.2$, and the Fermi energy $\ep_{\rm F} = 0.1$. 
As shown below, the obtained results are almost independent of the values of these coupling constants.
The upper critical field, where the magnetic field is perpendicular and parallel to the plane ($xy$ plane) in which the nodal-line resides, is denoted by $H_{c2,z}^{sym}$ and $H_{c2,x}^{sym}$. 
\par
In the same manner, Fig. \ref{fig:hc2_asym} shows the temperature dependence of the upper critical field for the asymmetric model with $a_{2}/a_{1} = 3.0$, $b/a_{1} = 1.0$, and $m = 3.0$ in the dirty limit (dotted lines) and the clean limit (dashed lines).
\par
There are several characteristic behaviors of $H_{c2}$.
\par
(1) The critical field is larger when the direction of the external magnetic field is in the $x$ direction, i.e., when the magnetic field is parallel to the nodal-line (orange lines). This trend is common for both the clean limit and the dirty limit. On the other hand, in the asymmetric model, the magnitude of the anisotropy is approximately three times larger than in the symmetric model. These behaviors are mainly due to the anisotropy of the Fermi velocity, as discussed later.
\par
(2) The temperature dependence of $H_{c2,z}$ in the clean limit is different from the other cases. At low temperatures $H_{c2,z}$ varies linearly, which is unusual in the s-wave superconductivity. 
\par
(3) In the asymmetric case (Fig. \ref{fig:hc2_asym}), $H_{c2}$ has a convex downward region, which does not appear in the symmetric model. In other words, $H_{c2}$ increases slowly with decreasing temperature. This is a property common to both clean and dirty limits.
\par
Note that the above results are for the case with $\ep_{\rm F}/a_{1} = 0.1$. Although we do not show the results, we find that the behaviors of $H_{c2}$ do not change unless $\ep_{\rm F}$ changes significantly.
We think that the above unconventional properties are mostly originating from the shape of the Fermi surface of the nodal-line as in Fig. \ref{fig:dispersion}, so the band crossing is not necessarily required. Furthermore, our numerical calculations show that the magnitude of coupling between branches has little effect on the qualitative behavior. In fact, changing the ratio $\lambda_{AB}/\lambda_{AA}$ or $\lambda_{BA}/\lambda_{AA}$ from 0 to 1 changes the upper critical field by less than 10 percent. From this, we conclude that the qualitative behavior remains unchanged even if we change the coupling constant.
\par
\begin{figure}[htbp]
    \begin{center}
        \includegraphics[keepaspectratio, width=1.0\linewidth]{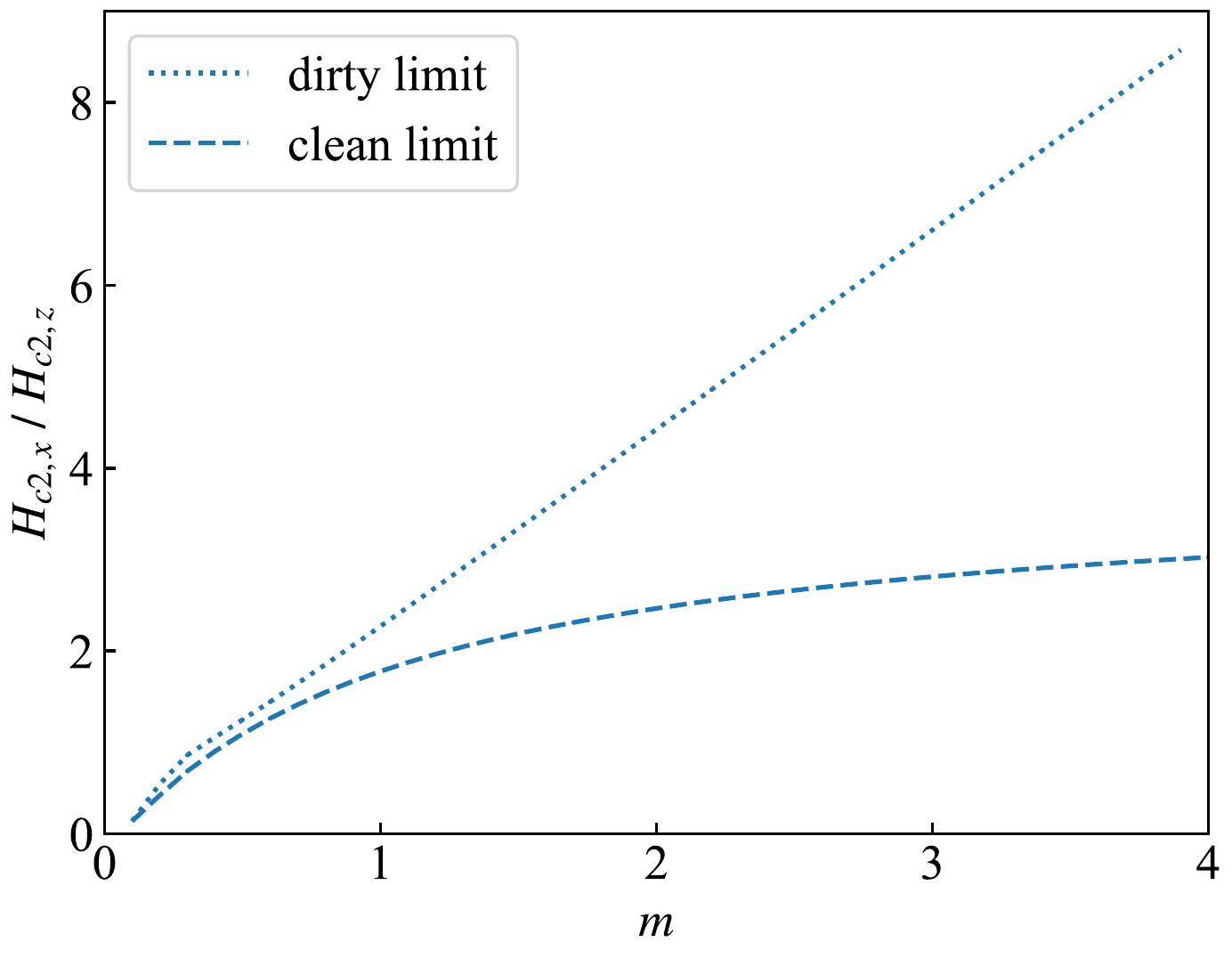}
        \caption{The anisotropy of the upper critical field of the asymmetric model $H^{asym}_{c2,x}/H^{asym}_{c2,z}$ at $T = 0$.}
        \label{fig:mplot}
    \end{center}
\end{figure}
\par
\section{Analysis of the behaviors of $H_{c2}$ and Comparison with the experiments}
\label{sec:analysis}
In what follows, we identify the origins of features (1)-(3), respectively.
\par
(1) The anisotropy of the upper critical field $H_{c2,x}/H_{c2,z}$ is proportional to the ratio $\sqrt{\mean{v_{x}^{2}}/\mean{v_{z}^{2}}}$. In fact, the anisotropy of the Fermi velocity is 
$\sqrt{\mean{v_{x}^{2}}/\mean{v_{z}^{2}}} \simeq 1.66$ for the symmetric model. This magnitude corresponds to the magnitude of the anisotropy of $H_{c2}$ as shown in Fig. \ref{fig:hc2_sym}. As shown in Figure \ref{fig:hc2_asym}, the anisotropy is more pronounced in the asymmetric model. This will be because $m$ is three times larger than the symmetric model. As $m$ increases, the Fermi velocity in the $x$ direction increases, and consequently, the critical field in the $x$ direction also increases. In fact, $\sqrt{\mean{v_{x}^{2}}/\mean{v_{z}^{2}}} \simeq 7.12$ for the asymmetric model. To see this more in detail, we show $m$ dependence of the ratio $H_{c2,x}^{asym}/H_{c2,z}^{asym}$ in Figure \ref{fig:mplot}.
We find that as $m$ increases, the ratio $H_{c2,x}^{asym}/H_{c2,z}^{asym}$ also increases, irrespective of the dirty and clean limit. Thus, the anisotropy of the critical field or the Fermi surface is mainly determined by the parameter $m$. It increases linearly with $m$ in the dirty limit, while it increases gradually in the clean limit. This will be due to the fact that $H_{c2,z}$ in the clean limit is larger than $H_{c2,z}$ in the dirty limit in the low temperature region because of the linear temperature dependence of $H_{c2,z}$ as discussed in (2) below.
\par
(2) The linear-$T$ dependence of $H_{c2,z}$ in the clean limit with the symmetric model (blue dashed line in Fig. \ref{fig:hc2_sym}) is understood as follows.
In the s-wave superconductivity, we can prove that $\dd H_{c2} / \dd T = 0$ at $T \to 0$ in generic cases. This holds when we can assume $\Lambda T / v_{\perp}$ is small, where $\Lambda = \pqty{2\abs{e}B}^{-1/2}$ and $v_{\perp}$ is the Fermi velocity perpendicular to the magnetic field. In a simple spherical Fermi surface, there are only two points where $v_{\perp}$ is zero. In the average over the Fermi surface, these points do not affect the result $\dd H_{c2} / \dd T |_{T=0} = 0$. However, in the present nodal-line model, when the magnetic field is applied along the $z$ direction, the set of points on the Fermi surface where $v_{\perp}$ is zero forms lines parallel to the nodal line. Therefore, the temperature region, in which the assumption that $\Lambda T / v_{\perp}$ is sufficiently small does not hold, is wider than in a simple Fermi surface. As a result, the critical field shows a linear temperature dependence down to low temperatures (see some details in Appendix).
On the other hand, the linear temperature dependence does not appear in the dirty limit. 
This difference may be due to the fact that the anisotropy is weakened by impurity scattering in the dirty limit, which makes it difficult to see the anisotropy that appears in the clean limit.
In the asymmetric model (Fig. \ref{fig:hc2_asym}), similar behaviors are observed. However, it seems that the linear temperature region in $H_{c2,x}$ is larger than in the symmetric case. This will be because the curvature of the Fermi surface increases with the increase of $m$, and the region of small $v_{\perp}$ becomes larger.
\par
(3) We also found that the introduction of asymmetry with respect to the nodal-line plane changes the behavior of the critical field near the transition temperature. This could be explained by the difference in Fermi velocities between the branches. 
Indeed, numerical calculations with dirty limit show that the value of $\dd^{2} H_{c2}/\dd T^{2}$ at $T = T_{c}$ is negative when $0.63 < D^{A}/D^{B} < 1.6$ and positive otherwise. In the symmetry model, $D^{A}/D^{B} \simeq 0.81$ (when the magnetic field is in the $z$ direction) and $D^{A}/D^{B} \simeq 0.91$ (when the magnetic field is in the $x$ direction) for the symmetric model, $D^{A}/D^{B} \simeq 8.87$ ($z$ direction) and $D^{A}/D^{B} \simeq 3.97$ ($x$ direction) for the asymmetric model. Thus, the difference in the Fermi velocity between the branches produces $H_{c2}$ that is convex downward.
\par
Experimental results of PbTaSe$_2$\cite{PhysRevB.93.054520} are also shown in the figure \ref{fig:hc2_asym} (solid lines with circles), which we think are consistent with the present theoretical results in the following points.
\par
(1) The experimental data show a large anisotropy in $H_{c2}$.  Specifically, $H_{c2}$ is larger when the magnetic field is parallel to the nodal-line. Such anisotropy is consistent with the behavior of our asymmetric model.
\par
(2) Although there is no experimental data at low temperatures when the field is perpendicular to the nodal-line (i.e., $H_{c2,z}$, solid blue circles in Fig. \ref{fig:hc2_asym}), it seems that $H_{c2,z}$ is linear near $T=0$. This behavior is consistent with our model. For $H_{c2,x}$, the experimental data (solid orange circles in Fig. \ref{fig:hc2_asym}) also show the linear behavior at low temperatures, which differs from that predicted by the present theory. However, we think that this point can be understood in the framework of the present theory if we extend the theory as follows.
In this paper, we considered a situation in which the nodal-line is completely on a plane perpendicular to some reciprocal lattice vector ($z$-axis). 
However, in actual materials, there exists a tilt and the nodal-line lies on a plane not perfectly perpendicular to some reciprocal lattice vector.
Due to this tilt, even when the magnetic field is in the $x$ direction, the region where $v_{\perp}$ is zero is not zero-dimensional but one-dimensional.
We speculate that this causes the linearity of the upper critical field at low temperatures.
To fit the experimental results completely, numerical methods with a material-dependent model Hamiltonian will be necessary.
\par
(3) The experimental data have convex downward behavior near the transition temperature, which is unusual for $H_{c2}$. This tendency is consistent with the present theory.
\section{Conclusion}
In this paper, we analyzed the temperature and Fermi energy dependence of the upper critical field for a typical model of s-wave nodal-line superconductors using the method of semiclassical Green's functions. The above analysis was performed for two different limits, the dirty limit with many impurities and the clean limit with few impurities. 
As a result of the calculations, the following characteristics were found for the upper critical magnetic field: (1) the anisotropy in the direction of the magnetic field, (2) linear behavior of $H_{c2,z}$ at low temperatures, and (3) convex downward behavior near the critical temperature. The above behaviors are different from those of ordinal s-wave superconductors but are consistent with those of experimental data of nodal-line superconductors.
This suggests that the model used in this study, as well as the semiclassical Green's function, is useful in the analysis of superconductivity in nodal-line semimetals.

\section*{Acknowledgement}
J. Endo was supported by the Japan Society for the Promotion of Science through the Program for Leading Graduate Schools (MERIT). We are grateful to Prof. Hiroshi Yasuoka for fruitful discussions. This work is supported
by Grants-in-Aid for Scientific Research from the Japan Society for the Promotion of Science (Nos. JP19K03720,
JP18H01162, and JP18K03482).

\appendix
\section{About the linear dependence}
First, we show the proof of $\dd H_{c2} / \dd T = 0$ near $T = 0$ in the clean limit. Considering the total derivative of eq. \eqref{eq:Sce2}, we obtain 
\begin{equation}
    \label{eq:appendix}
    \sum_{\alpha} N^{\alpha}\mean{\frac{16\Lambda^{2}}{v_{\perp}^{2}}\frac{T}{T_{c}}\dd T + \pqty{-\frac{8\Lambda^{2}}{v_{\perp}^{2}}\frac{T^{2}}{T_{c}^{2}} + 3} \frac{\dd B}{B}}_{\vk_{\rm F}^{\alpha}} = 0.
\end{equation}
Here, assuming $\Lambda T / v_{\perp}$ is small, we performed a Taylor expansion in terms of $\Lambda T / v_{\perp}$. It is clear that with the limit of $T \to 0$, $\dd B$ must approach zero. Therefore, we can conclude $\dd H_{c2} / \dd T = 0$.
\par
However, when $v_{\perp}$ is small, even if the temperature is low enough, the condition that $\Lambda T / v_{\perp}$ is small does not hold. In such a case, the above calculation is not valid.
In fact, assuming $\Lambda T / v_{\perp}$ is large, we obtain $\dd H_{c2} / \dd T \simeq -16\pi^{2}T^{2}/7\zeta(3) \abs{e} v_{\perp}^{2}$. $v_{\perp}$ varies according to its position on the Fermi surface, and \eqref{eq:appendix} includes averaging over the Fermi surface. Therefore, this argument is not rigorous. However, we can at least conclude that the temperature region, in which the gradient $\dd H_{c2} / \dd T$ is small, becomes narrower when the region, in which $v_{\perp}$ is small becomes larger. This explains the linear-$T$ dependence of $H_{c2,z}$ near $T = 0$ in the clean limit.
\par

\bibliographystyle{apsrev4-2}
\bibliography{main}

\end{document}